\documentclass[useAMS,usenatbib,preprint2]{mn2e}
\usepackage{natbib}
\usepackage{parskip}
\usepackage{hyperref}
\usepackage{epstopdf}
\usepackage{color}
\usepackage{float}
\usepackage{gensymb}
\usepackage[usenames,dvipsnames]{xcolor}
\usepackage{epsfig}
\usepackage{graphicx}

\newcommand{\bi}{\begin{itemize}}
\newcommand{\ei}{\end{itemize}}
\newcommand{\beq}{\begin{equation}}
\newcommand{\eeq}{\end{equation}}
\newcommand{\bea}{\begin{eqnarray}}
\newcommand{\eea}{\end{eqnarray}}
\newcommand{\bqu}{\begin{quote}}
\newcommand{\equ}{\end{quote}}
\newcommand{\bctr}{\begin{center}}
\newcommand{\ectr}{\end{center}}
\newcommand{\bd}{\begin{description}}
\newcommand{\ed}{\end{description}}
\newcommand{\bdm}{\begin{displaymath}}
\newcommand{\edm}{\end{displaymath}}
\newcommand{\lsim}{\mbox{$\:\stackrel{<}{_{\sim}}\:$} }

\newcommand{\om}{\Omega_{\rm m}}

\newcommand{\oll}{\Omega_\Lambda}

\newcommand{\vp}{\ensuremath{v_{\rm p}}}
\newcommand{\vr}{\ensuremath{v_{\rm r}}}
\newcommand{\vt}{\ensuremath{v_{\rm t}}}
\newcommand{\vrr}{\ensuremath{\bar{v}}}
\newcommand{\zp}{\ensuremath{z_{\rm p}}}

\newcommand{\zbar}{\ensuremath{\bar{z}}}
\newcommand{\vbar}{\ensuremath{\bar{v}}}
\newcommand{\kms}{\ensuremath{{\rm km\, s}^{-1}}} 
\newcommand{\kmsMpc}{\ensuremath{{\rm km\, s}^{-1}{\rm Mpc}^{-1}}} 

\def \aj{AJ}

\def \apj{ApJ}
\def \apjl{ApJ}
\def \apjl{ApJ Lett.}
\def \apjs{ApJS}

\def \araa{ARA\&A}

\def \jcap{J. Cosmo. Astropart. Phys.}

\def \mnras{MNRAS}

\def \pasp{PASP}

\def \physrep{Physics Reports}

\def \prd{Phys. Rev. D}

\title[Deriving peculiar velocities]{Deriving accurate peculiar velocities (even at high redshift)}

\author[Davis \& Scrimgeour]{
Tamara M.~Davis$^{1}$,\thanks{E-mail: tamarad@physics.uq.edu.au} and
Morag I.~Scrimgeour$^{2,3}$\\
$^{1}$School of Mathematics and Physics, University of Queensland, QLD 4072, Australia\\
$^{2}$Department of Physics and Astronomy, University of Waterloo, Waterloo, ON N2L 3G1, Canada\\
$^{3}$Perimeter Institute for Theoretical Physics, 31 Caroline Street North, Waterloo, ON N2L 2Y5, Canada
}

\begin{document}

\date{Accepted 2014 May 5. Received 2014 May 1; in original form 2014 April 1}

\pubyear{2014}

\maketitle

\label{firstpage}

\begin{abstract}
The way that peculiar velocities are often inferred from measurements of distances and redshifts makes an approximation, $\vp = cz-H_0D$, that gives significant errors even at relatively low redshifts (overestimates by $\Delta\vp\sim$100 \kms\ at $z\sim0.04$).  Here we demonstrate where the approximation breaks down, the systematic offset it introduces, and how the exact calculation should be implemented. \\
\end{abstract}

\begin{keywords}
cosmology -- peculiar velocities.
\end{keywords}


\section{Introduction}
Peculiar velocities can be a useful cosmological probe, as they are sensitive to the matter distribution on large scales, and can test the link between gravity and matter.  As peculiar velocity measurements are getting more numerous, more accurate, and stretching to higher redshift, it is important that we examine the assumptions made in the derivation of peculiar velocity from measurements of redshift and distance.  Some of the derivations used in the past have used approximations that are inadequate at the redshifts modern telescopes are able to reach. 

The starting point of many peculiar velocity papers is the equation, 
\beq  \vp = cz-H_0D,\label{eq:start} \eeq
where radial peculiar velocity $\vp$ is the difference between $cz$, the observed redshift $z$, multiplied by the speed of light, $c$; and the velocity given by Hubble's law, $\vbar=H_0 D$, where $H_0$ is Hubble's constant and $D$ is the proper distance. 

This appears in the classic papers such as \citet[][Eq.~1.1]{kaiser88}, \citet[Eq.~21]{dekel94}, and \citet[Eq.~1, 2, \& 147]{strauss95}, through to more recent peculiar velocity papers such as \citet[Eq.~5]{masters06}, \citet[Eq.~1]{springob07}, \citet[Eq.~1]{sarkar07}, \citet[Sect.~1, who note it is low-$z$ only]{abatebridle08}, \citet[Eq.~1]{lavaux08}, \citet[Eq.~5 \& 10]{nusserdavis11},\footnote{
Private communication with some groups indicates that although this equation appears in this paper they do not use it in their code \citep[e.g.][]{turnbull12}; others \citep[e.g.][]{abate09} do not specify how they derive velocities, but use data sets that have used the approximation; and yet others \citep[e.g. the COMPOSITE sample of ][]{watkins09,feldman10} use a combination of data sets that have used the approximation \citep[e.g.][]{springob07,springob09} with other data sets in which they have converted distances to velocities themselves without using the approximation.
} and is used in major compilations of peculiar velocity data such as Cosmicflows-2 \citep{tully13}.

This formula contains the approximation that, $v^{\rm approx} = cz$, which fails at high redshift. This has been pointed out in the past \citep[e.g.][]{faber77,harrison74,lynden-bell88,harrison93,colless01}, but since many recent papers still use the approximation, and because several major peculiar velocity surveys are imminent, it is timely to revisit this issue.  Most of the analyses listed above used data at low enough redshift that this approximation gives only small biases, but that will not remain true for future surveys.\footnote{For example, the  TAIPAN survey soon to start on the UK Schmidt telescope in Australia will measure distances to $\sim45,000$ elliptical galaxies out to $z\sim0.1$ \citep{koda14}; supernova surveys such as SkyMapper \citep{keller07} and Palomar Transient Factory \citep{law09,rau09} will be finding hundreds of supernovae in wide fields out to $z\sim0.1$ and $z\sim0.25$ respectively; and Tully-Fisher measurements from radio surveys such as WALLABY on the Australian Square Kilometre Array Pathfinder (ASKAP) will deliver $\sim32,000$ galaxy distances out to $z\sim0.1$ \citep{duffy12,koda14}.  At $z\sim0.1$ the approximation overestimates peculiar velocities by $\sim700\kms$.}   
In what follows we will demonstrate how the approximation causes one to overestimate peculiar velocities, and how that should be corrected for.

\section{The Basics}

Here we base our calculations in the Robertson-Walker metric, in which distance is given by $D=R\chi$, where $R(t)$ is the scalefactor at time $t$, denoted $R_0$ at the present day (dimensions of distance), and $\chi(\zbar)$ is the comoving coordinate of an object at cosmological redshift $\zbar$.  (Throughout we use overbars to denote quantities that would be measured in a perfectly homogeneous and isotropic universe without peculiar velocities.) 

Differentiating distance with respect to time (denoted by an overdot) gives the total velocity $v = \dot{R}\chi+R\dot{\chi}$.  Thus it is convenient to separate motion into recession velocities, due only to the Hubble flow, $\vrr = \dot{R}\chi$, and peculiar velocities, $\vp=R\,\dot{\chi}$, which encapsulate all motion other than the homogeneous and isotropic expansion, so 
\beq v = \vrr+\vp. \label{eq:v}\eeq

Peculiar velocities can be distinguished from recession velocities because an observer with a non-zero peculiar velocity sees a dipole in the cosmic microwave background. 
The comoving distance, $\chi$, to a galaxy is related to its cosmological redshift, $\zbar$, by
\beq \chi(\zbar) = \frac{c}{R_0}\int_0^{\zbar}\frac{dz}{H(z)}, \label{eq:chi} \eeq
where $c$ is the speed of light, and $H(\zbar)$ is the Hubble parameter as a function of cosmological redshift (assuming an homogeneous universe and smooth expansion, about which peculiar velocities are a small perturbation).
This distance is the cornerstone of any of the distances we measure using our distance probes, such as type Ia supernovae, fundamental plane, and Tully-Fisher distances.  The present day values of proper distance, $D$, luminosity distance, $D_{\rm L}$, and angular diameter distance, $D_{\rm A}$, are all built from comoving distance according to, respectively,
\bea D\; (\zbar)&=& R_0\;\chi(\zbar),\label{eq:D}\\
	D_{\rm L}(\zbar) &=& R_0\, S_{\rm k}(\chi)(1+\zbar), \\ 
         D_{\rm A}(\zbar) &=& R_0\, S_{\rm k}(\chi)(1+\zbar)^{-1},\eea
where $S_{\rm k}(\chi) = \sin(\chi), \chi, \sinh(\chi)$ for closed, flat, and open universes, respectively. 

When an object has a peculiar velocity, it acquires an additional redshift component $\zp$.  
The relationship between the peculiar velocity, and the `peculiar' redshift, \zp, is
\bea \vp=c {\zp} \label{eq:vpnonrel}, \eea
when the velocities are non-relativistic, or 
\beq \vp = c\frac{(1+\zp)^2-1}{(1+\zp)^2+1} , \label{eq:vprel}\eeq
when the velocities are relativistic.\footnote{Eq.~\ref{eq:vprel} is only strictly true if the velocities are entirely radial.  Converting between reference frames in special relativity depends on the total velocity.  So when there is also a tangential component $v_{\rm t}$, such that the peculiar velocity is broken up into $\vp^2 = \vr^2+\vt^2$, the relationship becomes: \newline $\vr = c\frac{-1+(1+\zp)\sqrt{(1+\zp)^2(1-(\vt/c)^2)-(\vt/c)^2}}{1+(1+\zp)^2}$.    \label{foot:vpecrel}}  In almost all practical situations the non-relativistic approximation is adequate for peculiar velocities.  The jets being ejected from active galactic nuclei would be one exception; cosmic rays would be another.

A galaxy with $\zbar$ and $\zp$ will appear to the observer to have $z$ where,
\beq (1+z) = (1+\zbar)(1+\zp). \label{eq:zsum} \eeq
Note that the approximation $z=\zbar + \zp$ works only for small redshifts.\footnote{Note also that the NASA/IPAC Extragalactic Database (NED) provides a velocity calculator that correctly uses Eq.~\ref{eq:v}, but if the user calculated $\vbar=c\zbar$, NED will give a biased result (see Sect.~\ref{sect:discussion}).}  Although Eq.~\ref{eq:zsum} is a standard result there is some confusion over this in the community so we run through the derivation in Appendix~\ref{app:A}.

\vspace{-5mm}
\section{The usual approximation and where it fails}

To measure a peculiar velocity one needs: 
\begin{itemize}
\item the observed redshift, $z$,
\item the observed distance, usually $D_{\rm L}$ or $D_{\rm A}$, from which one can infer $\chi(\zbar)$, and thus $\zbar$ (given a cosmological model).
\end{itemize}  

With $z$ and $\zbar$ known, Eq.~\ref{eq:zsum} gives $\zp$, from which peculiar velocity can be inferred.

Sticking with redshifts for this calculation avoids some of the approximations below, but conceptually, researchers have tended to prefer to work in velocities.  So the technique often used goes as follows:
\begin{itemize}
\item The observed redshift is used to infer the total velocity, $v$, which includes the recession velocity due to expansion of the universe $\vrr$, and the peculiar velocity $\vp$.   
\item The distance measurement is used to calculate the recession velocity as per Hubble's Law $\vrr = H_0 D$.  
\end{itemize}
The difference between these two velocities is the peculiar velocity (and measurement error), $\vp = v-\vrr$.

\begin{figure}
\includegraphics[width=84mm]{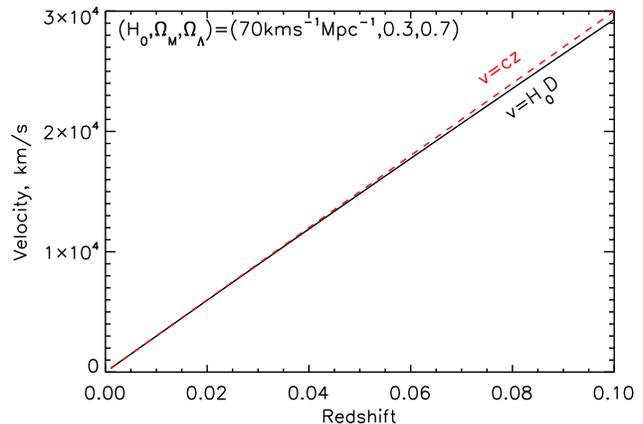}
\caption{The solid line shows recession velocity as a function of redshift in a standard model universe ($\Lambda$CDM) with matter density $\om=0.3$ and cosmological constant $\oll=0.7$.  The dashed line shows the approximation of $\vrr = c\zbar$.  Despite the fact that the deviation between these two looks small over this $z<0.1$ redshift range, the magnitude of the velocities in question is large, and therefore even this apparently small deviation is larger than the peculiar velocities we are interested in measuring.}
\label{fig:vr}
\end{figure}

However, calculating total velocity from observed redshift is tricky (as it needs knowledge of the peculiar velocity, which is what we are trying to measure).  So in the literature it is commonly approximated by $v^{\rm approx} = cz$, and thus  
\bea \vp^{\rm approx} &=& v^{\rm approx} -\vrr \\
  \vp^{\rm approx} &=& cz-H_0D.\label{eq:approx}
  \eea
However, the relationship between redshift and recession velocity does not follow that form, $\vrr\ne c\zbar$.  Fig.~\ref{fig:vr} shows that deviations are visible even at $z\lsim0.05$, and that the approximation causes one to overestimate peculiar velocities.

Note that all the velocities that appear in Eq.~\ref{eq:v} should be evaluated at the same cosmic time. Since we are measuring the peculiar velocity at the time of emission, $t_{\rm e}$, we should  be using the total and recession velocities at the time of emission too.  However, using $\vbar=H(t_{\rm e})D(t_e)$ actually makes the approximation in Eq.~\ref{eq:approx} even worse.   As you will see, we recommend always working in redshifts, never velocities, so this issue does not arise.



\vspace{-5mm}
\section{How bad is the approximation?}

\begin{figure}
\includegraphics[width=84mm]{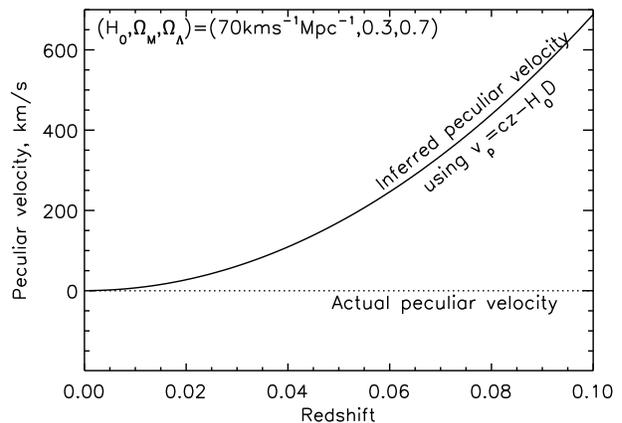}
\caption{The solid line shows the peculiar velocity that we would infer using the approximation in Eq.~\ref{eq:approx}, when the actual peculiar velocity is 0\,\kms.  The redshift on the horizontal axis refers to the cosmological redshift due to recession only.  Note that even as close as $z=0.04$ this approximation causes one to over-estimate the peculiar velocity by about 100\kms. }
\label{fig:vp_meas}
\end{figure}

To assess the effect of the approximation on our estimates of peculiar velocities, we will calculate the peculiar velocity with and without this approximation.  Let us start with an array of recession redshifts, $\zbar$, and for each one calculate: 
\begin{itemize}
\item $D$ the proper distance corresponding to that redshift (using Eq.~\ref{eq:chi} \&~\ref{eq:D}) --- this is the distance we can infer from luminosity or angular diameter distances,\footnote{Modulo knowledge of the cosmological model -- i.e.\ matter and dark energy densities \& properties -- so $S_k(\chi)$ and $H(z)/H_0$ can be calculated.  At low redshifts ($z\ll0.1$) this tends towards $D=cz/H_0$, and no cosmological model is required, which is one reason that approximation is so popular.} and
\item $z$ the redshift we would observe if the galaxy had a peculiar velocity, $\vp$, (using Eq.~\ref{eq:vprel} \&~\ref{eq:zsum}); initially we will set $\vp=0\;\kms$.
\end{itemize}
Then we use Eq.~\ref{eq:approx} to calculate the $\vp^{\rm approx}$ that we would infer for that galaxy.  So in essence, this test is taking information we do not have (the intrinsic redshift and true peculiar velocity of the galaxy), calculating what we would observe (the observed redshift and distance), and then calculating what the inferred peculiar velocity would be if we used the approximation in Eq.~\ref{eq:approx}.  Comparing this to the true peculiar velocity that we inputted shows how good (or bad) the approximation is.  The results are shown in Fig.~\ref{fig:vp_meas}.

From Fig.~\ref{fig:vp_meas} it is clear that the approximation is a poor one, even as close as $z=0.04$, where we would over-estimate $\vp$ by about 100\kms. 

Luckily this is easily fixed.  One can simply use the measured $D$ to infer what the recession redshift, $\zbar$, should be, apply the equation you would derive from rearranging Eq.~\ref{eq:zsum}, 
\beq \zp^{\rm correct} = \frac{z-\zbar}{1+\zbar}, \label{eq:zcorrect} \eeq
and then substitute that into the equation for peculiar velocity (Eq.~\ref{eq:vpnonrel}), or using the relativistic form (Eq.~\ref{eq:vprel}) if the peculiar velocity is large.  This form of the equation has been used, for example by \citet[][Eq.~10]{colless01} for the EFAR peculiar velocity measurements. 

Unfortunately to infer $\zbar$ from $D$ we need to assume a cosmological model.  So deriving peculiar velocities becomes dependent on the underlying cosmology.  A peculiar velocity measurement, therefore, really needs to be done either by assuming the underlying cosmological model (at which point one can only ask self-consistency questions about that model) or by doing a simultaneous fit to peculiar velocities and cosmology.

\vspace{-5mm}
\section{Impact on measurements of $H_0$}
Hubble constant measurements are made by taking an ensemble of objects with distance measurements (such as Cepheids or supernovae), using their measured redshifts, $z_i$, to infer velocities for each of them, $v(z_i)$, then taking the ratio to infer the Hubble parameter, the mean of all such measurements giving the overall $H_0$, 
\beq H_0 = \left<\frac{v(z_i)}{D(z_i)}\right>.\eeq
Early treatments \citep[e.g.][]{freedman01} approximated numerator by $v=cz$, which leads to an overestimate of the Hubble parameter.\footnote{Also noted by \citet{riess09}, footnote 11.}  Many of the latest Hubble constant measurements\footnote{\citet{freedman09} do not specify how they calculate velocities, but are at a lower redshift than \citet{riess11} so would be less sensitive to the choice of velocity calculation.    \citet{riess09,riess11} never explicitly write this as a velocity, but their comparison of magnitudes directly to luminosity distance using this expansion effectively does the same thing.  Note too that \citet{tully13} omits the factor of $(1+z)^{-1}$ in Eq.~\ref{eq:taylor} because they divide by luminosity distance, not proper distance.} \citep{riess09,riess11,tully13} remedy that by using a Taylor expansion of luminosity distance in terms of the deceleration parameter, $q_0$, and jerk $j_0$,  which effectively gives,
\beq v(z) = \frac{c z}{1+z} \left[1+\frac{1}{2}(1-q_0)z - \frac{1}{6}(1-q_0-3q_0^2+j_0)z^2\right],\label{eq:taylor}\eeq
where $q_0=-0.55$ and $j_0=1.0$ for standard $\Lambda$CDM with $(\om,\oll)\sim(0.3,0.7)$.
This gives a very good approximation to the full expression for velocity, out to $z\sim1$.   The full equation is, $v(z) = c\int_0^z\frac{dz}{E(z)}$, where $E(z)=H(z)/H_0$ depends on cosmological parameters but not $H_0$.  

Neglecting the cosmology dependence and using $v(z)=cz$ would overestimate the Hubble parameter by $\Delta H_0\sim1\,\kmsMpc$ for a sample evenly distributed in redshift out to $z\sim1$, so it is very important to include the cosmology dependence of $v(z)$ when measuring $H_0$.  The fear that doing so renders the measurement of $H_0$ cosmology dependent is unfounded.  All remotely viable Friedmann-Robertson-Walker cosmological models are closer to the fiducial model mentioned above than they are to $v=cz$.  So no matter what the cosmological model, the full expression for velocity using any fiducial model will always be better than the linear approximation in $z$, and the cosmology dependence of the resulting $H_0$ measurement is weak.

\vspace{-5mm}
\section{Discussion and Conclusions}\label{sect:discussion}
Until recently, most peculiar velocity surveys have been performed at $z\lsim0.02$, for which $v^{\rm approx} = cz$ gives less than a 50\kms\ error (see Fig.~\ref{fig:vp_meas}), less than the measurement uncertainty (but still biased on the high side).  However, as current and future peculiar velocity surveys probe ever deeper, and become more accurate, using the full formula for deriving peculiar velocities will be crucial.  This is particularly true for supernova surveys as they have the most precise distance measurements and can reach to high redshift.  For example, \citet{dai11} use a second-order approximation in their measurement of a bulk flow using supernova peculiar velocities in two bins on either side of $z=0.05$, while \citet{colin11} fit for a cosmological model, before comparing the magnitude residuals,\footnote{\citet{colin11} still seem to measure their bulk flow by the residual in distance vs the observed $z$, rather than $\zbar$, which could potentially induce a small bias, but one that is well below the uncertainty of the measurement. 
} and \citet{rathaus13} use a first order expansion with a fiducial model to analyse the Union 2.1 supernova sample out to $z<0.2$. 
Analyses using the kinetic Sunyaev-Zeldovich effect, which measure peculiar velocities at $z\sim0.1$, \citep[e.g.][]{kashlinsky08} {\em should not} be susceptible to this effect (which would be $\sim700\kms$ at $z=0.1$) because the temperature variation measured reflects only the peculiar velocity of the high-redshift cluster, not the recession velocity.\footnote{However, \citet{kashlinsky08} compares their results to a cosmological model in their Fig.~1f, using $\vbar=cz$.  Although that part is subtly incorrect, it doesn't alter their conclusion that they measured overly-large peculiar velocities, even though that has since been disputed for other reasons \citep{keisler09}.}

\begin{figure}
\includegraphics[width=84mm]{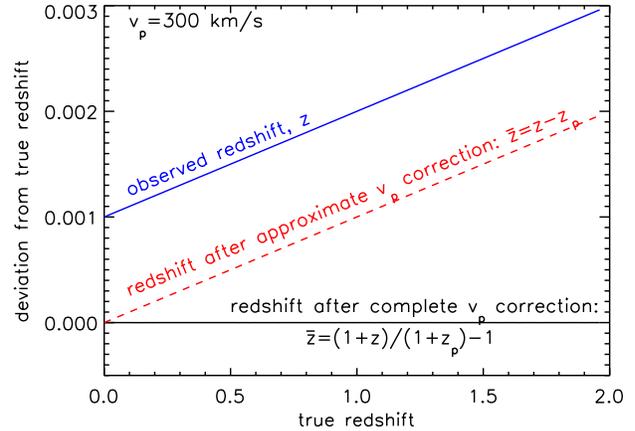}
\caption{When correcting for our own sun's motion of $\vp^{\rm obs}\sim300\kms$ with respect to the CMB frame, the errors due to the approximation are small.  This plot shows the maximum difference between the true cosmological redshift and (a) the observed redshift (blue); (b) the redshift after correcting using the approximation $\zbar = z-\zp$ (red dashed); and (c) the redshift after the complete correction has been applied $\zbar = (1+z)/(1+\zp)-1$.  Note that deviations are more significant at high redshift, where the magnitude-redshift diagram has a shallow slope, and therefore an incorrect $z$ has only a small effect on cosmology inferred from supernovae or large scale structure.  The calculations for this plot are done for the maximum peculiar velocity (when the direction of interest lies directly along our direction of motion with respect to the CMB). }
\label{fig:zcorr_us}
\end{figure}

Peculiar velocities can also affect the cosmological parameters estimated from the magnitude-redshift diagram of supernovae.  In \citet{davis11} we showed how to optimally de-weight low-redshift supernovae that potentially have correlated peculiar velocities that could bias cosmological measurements.  

The discussion in this paper is also relevant when correcting for our own motion.   Our sun moves at approximately $\sim300\kms$ with respect to the Hubble flow frame, as measured by our motion with respect to the cosmic microwave background (CMB).  If the $\zbar = z-\zp$ approximation is used when correcting observed redshifts into the CMB frame, then it introduces a systematic redshift error that increases with cosmological redshift.  Luckily that error will be small (see Fig.~\ref{fig:zcorr_us}).  We checked the potential effect on supernova cosmology results, for example, and it is negligible.

We also note here that peculiar velocities have a second-order effect, in that they perturb the observed magnitude in addition to the redshift \citep{hui06}.  In brief, the observed luminosity distance, $D_{\rm L}$, at the observed redshift, $z$, is related to the luminosity distance we would have seen in the absence of peculiar velocities, $\bar{D}_{\rm L}$, at the cosmological redshift $\zbar$, by 
\beq D_{\rm L}(z) = \bar{D}(\zbar)(1+\zp^{\rm obs})(1+\zp)^2,\eeq
where the two factors of $(1+\zp)$ come from Doppler shifting and relativistic beaming, and $\zp^{\rm obs}$ is the redshift due to our own motion along the line of sight direction.\footnote{\label{foot:zobs} The redshift due to our own motion arises only from the component of our motion along that line of sight.  When non-relativistic, $\zp^{\rm obs}=\vp^{\rm obs} \cos\theta/c$, where $\theta$ is the angle between our direction of motion and the line of sight.}  Since there is no factor of $(1+\zbar)$ in that equation, the correction is almost always small -- on the order of 0.3\% for peculiar velocities of both observer and emitter of $300\kms$.   We refer readers to \citet[Eq.~18]{davis11} for additional details of how to take that effect into account.  

Note that it is also important to use proper distance, not luminosity distance, when deriving velocities (this may have been done incorrectly in some early data sets).  Using luminosity distance would result in an offset from the true velocity of $\Delta v = \bar{v}z$, i.e. much larger than any error from the $v=cz$ approximation (about 3000\kms\ at $z=0.1$).

Another subtle aspect we may need to consider when analysing future large data sets, is that peculiar velocities tend to grow with cosmic time (at least until virialised).  Since we observe along a past light cone, we measure peculiar velocities at at a range of cosmic times.  Therefore the measured peculiar velocities will typically be lower than their velocities at the present day, which can cause problems if we average those velocities and try to measure a present-day bulk flow.  In order to correct peculiar velocities to the values they would take at the present day we would need to assume a model and use a value of $\om$, which is one of the quantities we are trying to measure.  So to compare velocity measurements to theory in detail, we should make our theoretical predictions along the past light cone. 

To conclude, we note that while it is tempting to apply a correction factor to the velocities published in papers that used the approximation, it is not quite that simple.  Peculiar velocity measurements are made complex by selection effects and calibration issues,\footnote{For example, the Tully-Fisher relation is often calibrated by minimising deviations from a linear Hubble law, $v=cz=HD$.  If that calibration is done at high enough distances for $v=cz$ to be inaccurate then that could introduce biases, but perhaps in a way that partially cancels the resulting error in peculiar velocities. \label{foot:caveat}} that need to be dealt with carefully for each data set.   Since bulk flow measurements are typically a conglomeration of distance measurements across a range of redshifts, the magnitude of the systematic error would be dependent on the redshift distribution of the catalogue being used; and if zero-point calibration has been done after the approximation has been made, this may ameliorate some of the bias the approximation could cause.

Moreover, it is not entirely clear that the approximation,  $\vp^{\rm approx}=cz-H_0 D$, that appears in so many papers has actually been applied in the codes that are used to calculate the peculiar velocities.  Equation~\ref{eq:zsum} is widely known, despite not being implemented in velocity calculators such as NED\footnote{http://ned.ipac.caltech.edu/forms/vel\_correction.html; Note that the equations in NED are correct, but they are prone to user error, if $\vbar=c\zbar$ is used to calculate the input.  When that occurs the magnitude of the error is small for most practical purposes (see Fig.~\ref{fig:zcorr_us}); even out at the CMB $z\sim1100$ the error would only be $\Delta z\sim 2$, or 0.2\%.}, and the fact that $cz=H_0D$ is an approximation is popular knowledge \citep[e.g.\ ][]{harrison74,harrison93}.  Private communication with some groups indicates that their code does not reflect the derivation given in their paper \citep[e.g.][]{turnbull12}.  So as alarming as a 100\kms\ error at $z=0.04$ would be (or a 700\kms\ error at $z=0.1$), some of the peculiar velocity analyses listed in the introduction may not suffer from much bias.  Nevertheless, it would be worth checking this point, especially in light of the fact that larger-than-expected peculiar velocities have been reported and claimed to challenge the standard $\Lambda$CDM cosmological model.  

Despite these caveats, we can give a rough indication of the magnitude of the correction that should be applied to some of the data sets, that we do know use the approximation, by considering the correction that would be required at their characteristic redshifts.  For example,
\begin{itemize}
\item SFI++ \citep{springob07} has median redshift  $\sim$0.02 with a maximum at $\sim$0.07, so may have overestimated velocities by a median of $\sim$20\kms\ and a maximum of $\sim$250\kms\ (modulo Footnote~\ref{foot:caveat}).
\item The Cosmicflows-2 compendium of distance measurements \citep{tully13}, when reanalysed with Eq.~\ref{eq:correct} finds a shift of 57\kms\ closer to a mean peculiar velocity of zero (the mean is still negative, but less so, now at -221\kms), as well as a small reduction in the dispersion of their sample on the order of 1\% (Tully, private communication). 
\end{itemize}

Meanwhile, the formula that should become the starting point for future peculiar velocity papers is (from Eq.~\ref{eq:zcorrect}), 
\beq \vp = c\left(\frac{z-\zbar}{1+\zbar}\right), \label{eq:correct}\eeq
where $\zbar$ is calculated from a distance measurement, using a fiducial cosmological model. 

\vspace{-5mm}
\section*{Acknowledgments}
We would like to thank the authors of many of the peculiar velocity papers mentioned here for checking their data sets in light of this discussion, including Alexandra Abate, Sarah Bridle, Hume Feldman, Chris Springob, Brent Tully, and Rick Watson, and in particular for extensive comments from Michael Hudson.
We would like to thank the 6dFGS peculiar velocities survey group for the discussions in our telecons that inspired this paper, and for feedback on the drafts, in particular Yin-Zhe Ma, Chris Blake, Matthew Colless, Andrew Johnson, John Lucey, Heath Jones, Jun Koda,  Christina Magoulas, Jeremy Mould, and Chris Springob; and also Eric Linder for pointing out footnote~\ref{foot:vpecrel}. 
TMD acknowledges the support of the Australian Research Council through a Future Fellowship award, FT100100595.  
MIS acknowledges the support of a Jean Rogerson Scholarship, a top-up scholarship from the University of Western Australia, and a CSIRO Malcolm McInstosh Lecture bankmecu scholarship.  
We both acknowledge the support of the ARC Centre of Excellence for All Sky Astrophysics, funded by grant CE110001020; and a UWA-UQ Bilateral Research Collaboration Award.


\begin{thebibliography}{35}
\expandafter\ifx\csname natexlab\endcsname\relax\def\natexlab#1{#1}\fi

\bibitem[{{Abate} {et~al}\mbox{.}(2008){Abate}, {Bridle}, {Teodoro}, {Warren},
  \& {Hendry}}]{abatebridle08}
{Abate} A., {Bridle} S., {Teodoro} L.~F.~A., {Warren} M.~S., {Hendry} M., 2008,
  \mnras, 389, 1739

\bibitem[{{Abate} \& {Erdo{\v g}du}(2009)}]{abate09}
{Abate} A., {Erdo{\v g}du} P., 2009, \mnras, 400, 1541

\bibitem[{{Carroll}(2004)}]{carroll04}
{Carroll} S.~M., 2004, {Spacetime and Geometry}. Addison Wesley, San Francisco

\bibitem[{{Colin} {et~al}\mbox{.}(2011){Colin}, {Mohayaee}, {Sarkar}, \&
  {Shafieloo}}]{colin11}
{Colin} J., {Mohayaee} R., {Sarkar} S., {Shafieloo} A., 2011, \mnras, 414, 264

\bibitem[{{Colless} {et~al}\mbox{.}(2001){Colless}, {Saglia}, {Burstein},
  {Davies}, {McMahan}, \& {Wegner}}]{colless01}
{Colless} M., {Saglia} R.~P., {Burstein} D., {Davies} R.~L., {McMahan} R.~K.,
  {Wegner} G., 2001, \mnras, 321, 277

\bibitem[{{Dai} {et~al}\mbox{.}(2011){Dai}, {Kinney}, \& {Stojkovic}}]{dai11}
{Dai} D.-C., {Kinney} W.~H., {Stojkovic} D., 2011, \jcap, 4, 15

\bibitem[{{Davis} {et~al}\mbox{.}(2011){Davis}, {Hui}, {Frieman},
  {Haugb{\o}lle}, {Kessler}, {Sinclair}, {Sollerman}, {Bassett}, {Marriner},
  {M{\"o}rtsell}, {Nichol}, {Richmond}, {Sako}, {Schneider}, \&
  {Smith}}]{davis11}
{Davis} T.~M. {et~al.}, 2011, \apj, 741, 67

\bibitem[{{Dekel}(1994)}]{dekel94}
{Dekel} A., 1994, \araa, 32, 371

\bibitem[{{Duffy} {et~al}\mbox{.}(2012){Duffy}, {Meyer}, {Staveley-Smith},
  {Bernyk}, {Croton}, {Koribalski}, {Gerstmann}, \& {Westerlund}}]{duffy12}
{Duffy} A.~R., {Meyer} M.~J., {Staveley-Smith} L., {Bernyk} M., {Croton} D.~J.,
  {Koribalski} B.~S., {Gerstmann} D., {Westerlund} S., 2012, \mnras, 426, 3385

\bibitem[{{Faber} \& {Dressler}(1977)}]{faber77}
{Faber} S.~M., {Dressler} A., 1977, \aj, 82, 187

\bibitem[{{Feldman} {et~al}\mbox{.}(2010){Feldman}, {Watkins}, \&
  {Hudson}}]{feldman10}
{Feldman} H.~A., {Watkins} R., {Hudson} M.~J., 2010, \mnras, 407, 2328

\bibitem[{{Freedman} {et~al}\mbox{.}(2009){Freedman}, {Burns}, {Phillips},
  {Wyatt}, {Persson}, {Madore}, {Contreras}, {Folatelli}, {Gonzalez}, {Hamuy},
  {Hsiao}, {Kelson}, {Morrell}, {Murphy}, {Roth}, {Stritzinger}, {Sturch},
  {Suntzeff}, {Astier}, {Balland}, {Bassett}, {Boldt}, {Carlberg}, {Conley},
  {Frieman}, {Garnavich}, {Guy}, {Hardin}, {Howell}, {Kessler}, {Lampeitl},
  {Marriner}, {Pain}, {Perrett}, {Regnault}, {Riess}, {Sako}, {Schneider},
  {Sullivan}, \& {Wood-Vasey}}]{freedman09}
{Freedman} W.~L. {et~al.}, 2009, \apj, 704, 1036

\bibitem[{{Freedman} {et~al}\mbox{.}(2001){Freedman}, {Madore}, {Gibson},
  {Ferrarese}, {Kelson}, {Sakai}, {Mould}, {Kennicutt}, {Ford}, {Graham},
  {Huchra}, {Hughes}, {Illingworth}, {Macri}, \& {Stetson}}]{freedman01}
{Freedman} W.~L. {et~al.}, 2001, \apj, 553, 47

\bibitem[{{Harrison}(1974)}]{harrison74}
{Harrison} E.~R., 1974, \apjl, 191, L51

\bibitem[{{Harrison}(1993)}]{harrison93}
{Harrison} E.~R., 1993, \apj, 403, 28

\bibitem[{{Hui} \& {Greene}(2006)}]{hui06}
{Hui} L., {Greene} P.~B., 2006, \prd, 73, 123526

\bibitem[{{Kaiser}(1988)}]{kaiser88}
{Kaiser} N., 1988, \mnras, 231, 149

\bibitem[{{Kashlinsky} {et~al}\mbox{.}(2008){Kashlinsky}, {Atrio-Barandela},
  {Kocevski}, \& {Ebeling}}]{kashlinsky08}
{Kashlinsky} A., {Atrio-Barandela} F., {Kocevski} D., {Ebeling} H., 2008,
  \apjl, 686, L49

\bibitem[{{Keisler}(2009)}]{keisler09}
{Keisler} R., 2009, \apjl, 707, L42

\bibitem[{{Keller} {et~al}\mbox{.}(2007){Keller}, {Schmidt}, {Bessell},
  {Conroy}, {Francis}, {Granlund}, {Kowald}, {Oates}, {Martin-Jones},
  {Preston}, {Tisserand}, {Vaccarella}, \& {Waterson}}]{keller07}
{Keller} S.~C. {et~al.}, 2007, Publications of the Astronomical Society of
  Australia, 24, 1

\bibitem[{{Koda} {et~al}\mbox{.}(2013){Koda}, {Blake}, {Davis}, {Magoulas},
  {Springob}, {Scrimgeour}, {Johnson}, {Poole}, \& {Staveley-Smith}}]{koda14}
{Koda} J. {et~al.}, 2014, arXiv:1312.1022

\bibitem[{{Lavaux} {et~al}\mbox{.}(2008){Lavaux}, {Mohayaee}, {Colombi},
  {Tully}, {Bernardeau}, \& {Silk}}]{lavaux08}
{Lavaux} G., {Mohayaee} R., {Colombi} S., {Tully} R.~B., {Bernardeau} F.,
  {Silk} J., 2008, \mnras, 383, 1292

\bibitem[{{Law} {et~al}\mbox{.}(2009){Law}, {Kulkarni}, {Dekany}, {et~al}}]{law09}
{Law} N.~M. {et~al.}, 2009, \pasp, 121, 1395

\bibitem[{{Lynden-Bell} {et~al}\mbox{.}(1988){Lynden-Bell}, {Faber},
  {Burstein}, {Davies}, {Dressler}, {Terlevich}, \& {Wegner}}]{lynden-bell88}
{Lynden-Bell} D., {Faber} S.~M., {Burstein} D., {Davies} R.~L., {Dressler} A.,
  {Terlevich} R.~J., {Wegner} G., 1988, \apj, 326, 19

\bibitem[{{Masters} {et~al}\mbox{.}(2006){Masters}, {Springob}, {Haynes}, \&
  {Giovanelli}}]{masters06}
{Masters} K.~L., {Springob} C.~M., {Haynes} M.~P., {Giovanelli} R., 2006, \apj,
  653, 861

\bibitem[{{Nusser} \& {Davis}(2011)}]{nusserdavis11}
{Nusser} A., {Davis} M., 2011, \apj, 736, 93

\bibitem[{{Rau} {et~al}\mbox{.}(2009){Rau}, {Kulkarni}, {Law}, {et~al}}]{rau09}
{Rau} A. {et~al.}, 2009, \pasp, 121, 1334

\bibitem[{{Rathaus} {et~al}\mbox{.}(2013){Rathaus}, {Kovetz}, \&
  {Itzhaki}}]{rathaus13}
{Rathaus} B., {Kovetz} E.~D., {Itzhaki} N., 2013, \mnras, 431, 3678

\bibitem[{{Riess} {et~al}\mbox{.}(2009){Riess}, {Macri}, {Casertano}, {Sosey},
  {Lampeitl}, {Ferguson}, {Filippenko}, {Jha}, {Li}, {Chornock}, \&
  {Sarkar}}]{riess09}
{Riess} A.~G. {et~al.}, 2009, \apj, 699, 539

\bibitem[{{Riess} {et~al}\mbox{.}(2011){Riess}, {Macri}, {Casertano},
  {Lampeitl}, {Ferguson}, {Filippenko}, {Jha}, {Li}, \& {Chornock}}]{riess11}
{Riess} A.~G. {et~al.}, 2011, \apj, 730, 119

\bibitem[{{Sarkar} {et~al}\mbox{.}(2007){Sarkar}, {Feldman}, \&
  {Watkins}}]{sarkar07}
{Sarkar} D., {Feldman} H.~A., {Watkins} R., 2007, \mnras, 375, 691

\bibitem[{{Springob} {et~al}\mbox{.}(2007){Springob}, {Masters}, {Haynes},
  {Giovanelli}, \& {Marinoni}}]{springob07}
{Springob} C.~M., {Masters} K.~L., {Haynes} M.~P., {Giovanelli} R., {Marinoni}
  C., 2007, \apjs, 172, 599

\bibitem[{{Springob} {et~al}\mbox{.}(2009){Springob}, {Masters}, {Haynes},
  {Giovanelli}, \& {Marinoni}}]{springob09}
{Springob} C.~M., {Masters} K.~L., {Haynes} M.~P., {Giovanelli} R., {Marinoni}
  C., 2009, \apjs, 182, 474

\bibitem[{{Strauss} \& {Willick}(1995)}]{strauss95}
{Strauss} M.~A., {Willick} J.~A., 1995, \physrep, 261, 271

\bibitem[{{Tully} {et~al}\mbox{.}(2013){Tully}, {Courtois}, {Dolphin},
  {Fisher}, {H{\'e}raudeau}, {Jacobs}, {Karachentsev}, {Makarov}, {Makarova},
  {Mitronova}, {Rizzi}, {Shaya}, {Sorce}, \& {Wu}}]{tully13}
{Tully} R.~B. {et~al.}, 2013, \aj, 146, 86

\bibitem[{{Turnbull} {et~al}\mbox{.}(2012){Turnbull}, {Hudson}, {Feldman},
  {Hicken}, {Kirshner}, \& {Watkins}}]{turnbull12}
{Turnbull} S.~J., {Hudson} M.~J., {Feldman} H.~A., {Hicken} M., {Kirshner}
  R.~P., {Watkins} R., 2012, \mnras, 420, 447

\bibitem[{{Watkins} {et~al}\mbox{.}(2009){Watkins}, {Feldman}, \&
  {Hudson}}]{watkins09}
{Watkins} R., {Feldman} H.~A., {Hudson} M.~J., 2009, \mnras, 392, 743

\end{thebibliography}

\appendix
\vspace{-5mm}
\section{Derivation of observed redshift}\label{app:A}\vspace{-3mm}
Abbreviating the emitted and observed wavelengths as $\lambda_{\rm e}$ and $\lambda_{\rm o}$, respectively, the definition of redshift is,
\beq z \equiv \frac{\lambda_{\rm o}-\lambda_{\rm e}}{\lambda_{\rm e}} \quad {\rm or}\,{\rm equivalently}\quad  1+z \equiv\frac{\lambda_{\rm o}}{\lambda_{\rm e}}. \eeq

\begin{figure}
\includegraphics[width=84mm]{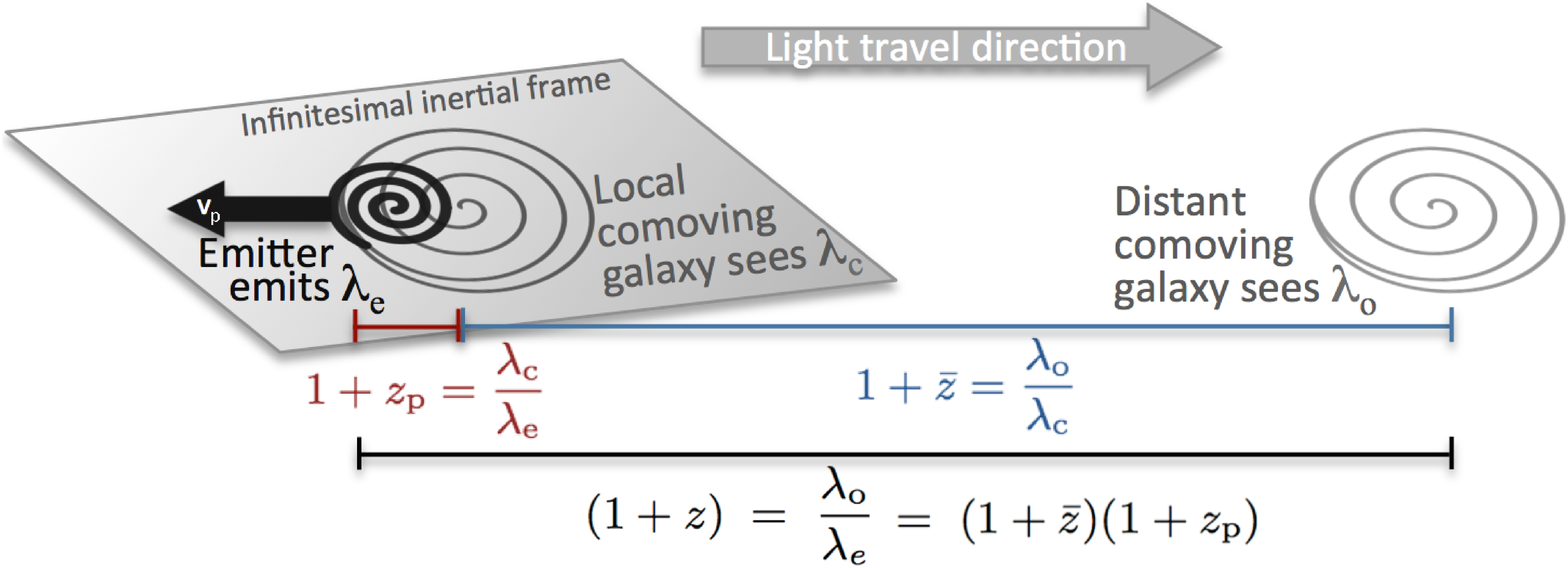}
\caption{The emitting galaxy, with peculiar velocity $v_{\rm p}$, can be considered to be instantaneously in an infinitesimal inertial frame with a hypothetical comoving observer at the same position.  Thus special relativity applies in converting $v_{\rm p}$ to $\zp$. (The separation represented by the red bar in this sketch is for illustrative purposes only, the emitter and local comoving observer should be coincident).  The photon then experiences the same cosmological redshift, $\zbar$, that any photon emitted from the local comoving galaxy would experience.   It is clear from the ratio of emitted and observed wavelengths that to get the total observed redshift, $z$, the cosmological redshift, $\zbar$, should be combined with the peculiar redshift, $\zp$, multiplicatively according to Eq.~\ref{eq:zmult}.}
\label{fig:app}
\end{figure}

Imagine you have three galaxies.  (1) An emitter with peculiar velocity, which emits light at wavelength $\lambda_{\rm e}$; (2) A local comoving galaxy (at the same position as the emitter), which sees the light from the emitter at $\lambda_{\rm c}$; and (3)  A distant comoving galaxy, which sees the light from the emitter at $\lambda_{\rm o}$.  (Assume the peculiar velocity is along the line of sight between the distant comoving galaxy and the emitter.  See Fig.~\ref{fig:app}.)

The redshift between the emitter and local comoving observer would be $1+\zp = \lambda_{\rm c}/\lambda_{\rm e}$.  This redshift occurs between two coincident observers (two observers at the same position, each with their own infinitesimal inertial frame), therefore special relativity applies, and the redshift is related to peculiar velocity by Eq.~\ref{eq:vpnonrel} or Eq.~\ref{eq:vprel}. \citep[For a technical discussion of infinitesimal inertial frames see][Sect.~2.5.]{carroll04}

The redshift between the local comoving observer and the distant comoving observer would be $1+\zbar = \lambda_{\rm o} / \lambda_{\rm c}$.   This is the cosmological redshift, and is the redshift that all photons would experience en route from that position.  It is related to the comoving distance by Eq.~\ref{eq:chi}.

So the total redshift between the emitter and the distant comoving observer would be,
\beq (1+z) \;=\; \frac{\lambda_{\rm o}}{\lambda_e} \;=\; \frac{\lambda_{\rm o}}{\lambda_c}\frac{\lambda_c}{\lambda_e} \;=\; (1+\zbar)(1+\zp).\label{eq:zmult} \eeq

That is the redshift that we would observe, if we were comoving.  If we also have a peculiar velocity of our own (which we do), that adds an extra peculiar redshift at the point of observation, $\zp^{\rm obs}$ (defined in footnote~\ref{foot:zobs}).  So following the same reasoning as above we find,
\bea (1+z) \;=\; (1+\zp^{\rm obs})(1+\zbar)(1+\zp). \eea

\end{document}